\documentclass{article}

    \PassOptionsToPackage{numbers, compress}{natbib}

 \usepackage[main, final]{neurips_2026}

\usepackage[utf8]{inputenc} %
\usepackage[T1]{fontenc}    %
\usepackage{hyperref}       %
\usepackage{url}            %
\usepackage{booktabs}       %
\usepackage{amsfonts}       %
\usepackage{nicefrac}       %
\usepackage{microtype}      %
\usepackage{xcolor}         %
\usepackage{graphicx}
\usepackage{float}
\usepackage{multirow}
\usepackage{amsmath}
\usepackage{pifont}%
\newcommand{\cmark}{\ding{51}}%
\newcommand{\xmark}{\ding{55}}%

\title{Your Embedding Model is SMARTer Than You Think}

\author{%
  Jianrui Zhang$^*$ \\
  UW-Madison \\
  \texttt{harrisz@cs.wisc.edu} \\
  \And
  Hyun Jung Lee\thanks{Equal Contribution} \\
  Korea University\\
  \texttt{hyulee@korea.ac.kr} \\
    \And
  Sukanta Ganguly \\
  NetApp, Inc. \\
  \texttt{sukanta.ganguly@netapp.com} \\
  \AND
  Tae-Eui Kam \\
  Korea University \\
  \texttt{kamte@korea.ac.kr} \\
  \And
  Donghyun Kim\thanks{Equal Advising} \\
  Korea University \\
  \texttt{d\_kim@korea.ac.kr} \\
  \And
  Yong Jae Lee$^\dagger$ \\
  UW-Madison \\
  \texttt{yongjaelee@cs.wisc.edu}
}

\begin{document}

\maketitle

\begin{figure} [h!]
    \centering
    \includegraphics[width=\linewidth]{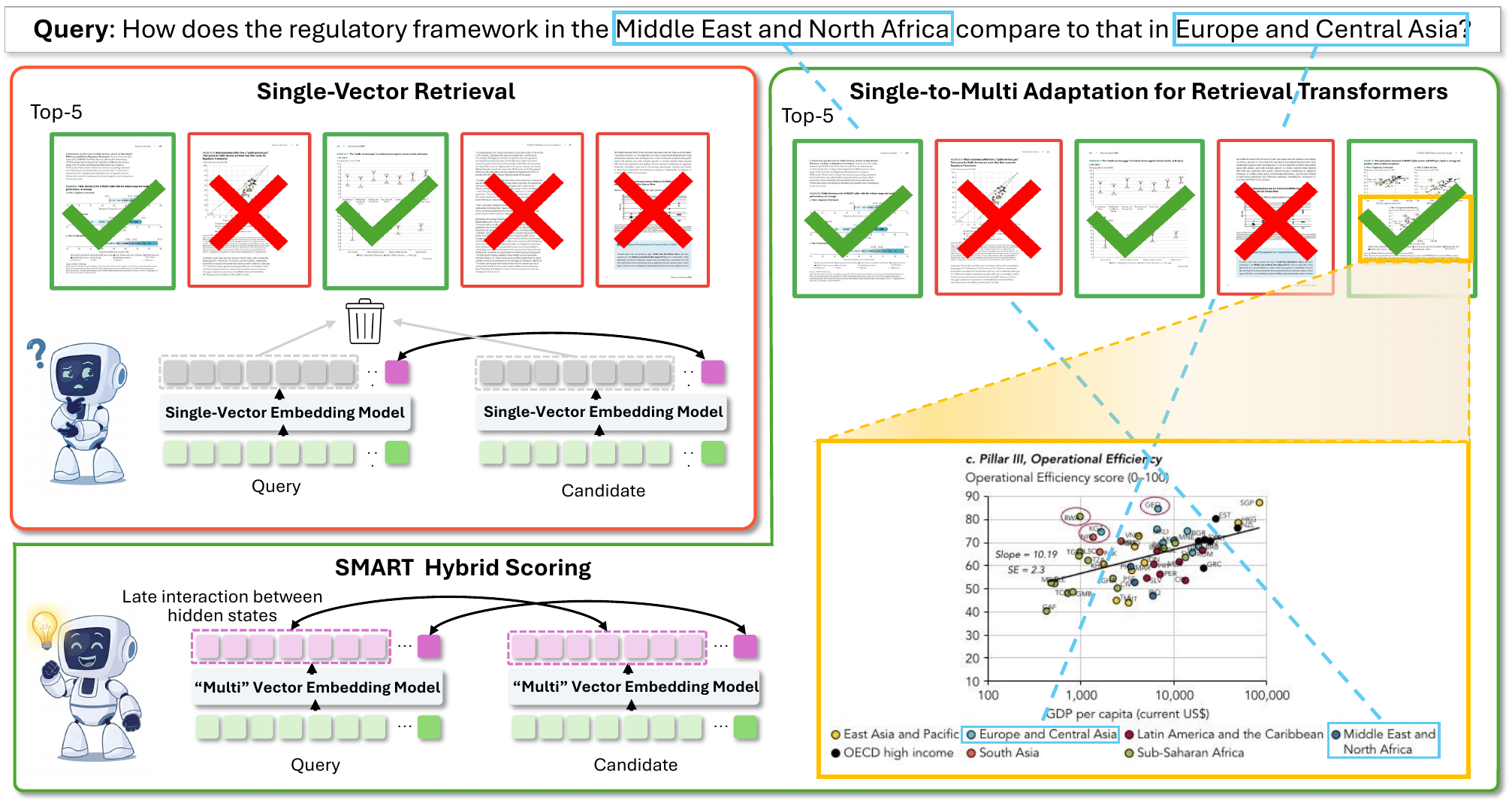}
    \caption{Standard single-vector embedding models (Qwen3-VL-Embedding-8B in this figure) compress sequences into single token representations, which often results in losing local information and underutilizing the remaining hidden states. 
    While existing multi-vector approaches require expensive retraining, SMART %
shows empirically that the hidden states of single-vector embedders are \textbf{already} geometrically aligned for local matching. SMART thus transforms any single-vector model into a multi-vector variant \textbf{both during inference and via lightweight post-training}, the former %
improving with no training, and the latter saving time and rivaling SoTA multi-vector embedders.}
    \label{fig:teaser}
\end{figure}

\begin{abstract}
    Multimodal retrieval relies heavily on single-vector retrievers, which compress rich, sequential token sequences into one single global representation. While efficient, they discard fine-grained, local evidence critical for dense retrieval tasks. Multi-vector approaches were introduced as a solution, but they strictly require training and many ignore the necessity of a globally summarizing representation. To address this, we introduce SMART, a framework that unlocks the latent multi-vector capabilities of standard single-vector models. %
We first observe that gradients from the pooled contrastive objective reach the preceding hidden states, and then show empirically that these states are already organized for cosine-based matching. By applying direct late-interaction over these frozen hidden states during inference, SMART acts as a plug-and-play upgrade that consistently improves performance across diverse modalities, improving even the state-of-the-art models further on MMEB-V2. %
We also show that lightweight post-training saves time and compute while further improving Visual Document retrieval, allowing a single-vector model to perform competitively with SoTA multi-vector counterparts. SMART targets dense retrieval: its gains concentrate on fine-grained and visual-document tasks, it is on par on global tasks such as classification, and it trades additional per-token index storage for these gains. Ultimately, SMART offers both a highly efficient inference enhancement and a powerful finetuning technique for multimodal retrieval.
    We open source our code and weights at \url{https://github.com/HanSolo9682/SMART}.
\end{abstract}

\section{Introduction}

Multimodal Large Language Models (MLLMs) have recently unified dense retrieval across text, images, visual documents, and videos~\cite{wei2023uniirtrainingbenchmarkinguniversal, jiang2025vlm2vectrainingvisionlanguagemodels}. State-of-the-art (SoTA) systems, such as the Qwen3-VL-Embedding series~\cite{li2026qwen3vlembeddingqwen3vlrerankerunifiedframework}, map these diverse modalities into a highly expressive, shared representation space, enabling efficient global similarity matching. However, these architectures predominantly rely on a single-vector paradigm, collapsing the entire sequence of multimodal hidden states into a single pooling token, such as the end-of-text (\texttt{<eot>}) token. While this compression ensures highly efficient indexing and nearest-neighbor search, recent theoretical analyses demonstrate fundamental limitations in its capacity~\cite{weller2026theoreticallimitationsembeddingbasedretrieval, luan2021sparsedenseandattentionalrepresentations, reimers2021curseofdenselow-dimensional}. Since the number of distinct subset rankings a single-vector paradigm can reliably return is strictly bounded by the embedding dimensionality, fine-grained multimodal queries that depend on fine-grained details such as local text, specific visual attributes, or regional bindings can often fail. This is because localized evidence encoded by the transformer can be lost when the input is compressed into the pooled representation used for scoring.

To overcome this expressive bottleneck, researchers have increasingly turned to multi-vector architectures, pioneered in the text domain by ColBERT~\cite{khattab2020colbertefficienteffectivepassage} and recently adapted for multimodal tasks via models like Colpali~\cite{faysse2025colpali} and jina-embeddings-v4~\cite{gunther2025jinaembeddingsv4universalembeddingsmultimodal}. Yet, these approaches require either full-scale task-specific finetuning or the introduction of learnable tokens (e.g., MetaEmbed~\cite{xiao2026metaembedscalingmultimodalretrieval}), which incurs significant computational and memory costs during training as they scale quadratically with respect to sequence length. 
Moreover, methods like Colpali and jina emphasize local token- or patch-level matching without explicitly preserving the global pooled readout that single-vector models use effectively.

To bridge this gap, we introduce SMART (\textbf{S}ingle-to-\textbf{M}ulti \textbf{A}daptation for \textbf{R}etrieval \textbf{T}ransformers), a framework that converts a single-vector retriever into a multi-vector retriever possible both at inference time and via lightweight finetuning, while preserving its global compatibility signal. %
We show empirically that the non-pooling hidden states of single-vector embedders are directly usable for cosine-based retrieval, even though contrastive supervision is applied only to the pooled representation. The local semantic evidence in these states can thus be utilized through late interaction without additional training. SMART initially exploits this by applying an additional late-interaction mechanism (\textsc{MaxSim})~\cite{khattab2020colbertefficienteffectivepassage} over the pre-pooling hidden states at inference time, combined with the pooled score in a hybrid scoring scheme, effectively recovering localized details without the overhead of training a multi-vector retriever from scratch. Building on this foundation, we further show that lightweight finetuning under the SMART objective yields additional gains, transforming single-vector embedders into competitive multi-vector retrievers. Critically, this conversion saves at least 20\% of training time and computation compared to training a multi-vector retriever from scratch under the same recipe. SMART thus unlocks the multi-vector capabilities of existing single-vector retrievers, improving retrieval without additional training and enabling further gains through efficient post-training.

Our core contributions are as follows:
\begin{itemize}
    \item We show that single-vector retrievers, despite being trained only with a pooled contrastive objective, \textbf{already} retain localized semantic evidence in their non-pooling hidden states. This makes it possible to convert an existing single-vector model into a multi-vector retriever by reusing these hidden states for token-level matching.
    \item We propose SMART, which can act as a \textbf{training-free, plug-and-play} upgrade using our \textbf{hybrid scoring} technique. It steadily improves retrieval accuracy across various complex retrieval tasks and backbones, %
including the SoTA Qwen3-VL-Embedding~\cite{li2026qwen3vlembeddingqwen3vlrerankerunifiedframework} %
series%
.
    \item We demonstrate that SMART can be further improved with \textbf{efficient post-training}, either by attaching a lightweight projection adapter while freezing the pretrained single-vector backbone, or by finetuning the single-vector embedder with our hybrid scoring objective. These variants enable single-vector embedders to achieve strong multi-vector retrieval performance, saving the time to train a dedicated multi-vector model from scratch. 
    
\end{itemize}

\section{Related Work}

\paragraph{Single-Vector Embedding Models}
Early work on contrastive models (CLIP~\cite{radford2021learningtransferablevisualmodels}, BLIP~\cite{li2022blipbootstrappinglanguageimagepretraining}, SigLIP~\cite{zhai2023sigmoidlosslanguageimage}) and MLLMs~\cite{liu2023llava,zhu2023minigpt4,liu2023improvedllava,Qwen-VL} paved the way for modern MLLM-based dense retrievers like UniIR~\cite{wei2023uniirtrainingbenchmarkinguniversal} and VLM2Vec~\cite{jiang2025vlm2vectrainingvisionlanguagemodels}. Recent advances focus on efficient training strategies (E5-V~\cite{jiang2024e5vuniversalembeddingsmultimodal}, GME~\cite{zhang2025gmeimprovinguniversalmultimodal}) and highly expressive unified spaces, with Qwen3-VL-Embedding~\cite{li2026qwen3vlembeddingqwen3vlrerankerunifiedframework} currently achieving SoTA. These approaches, however, %
discard the local fine-grained non-pooled hidden states, which SMART reuses to convert off-the-shelf single-vector models into multi-vector retrievers without training. %

\paragraph{Multi-Vector %
 and Late-Interaction Retrieval}
Because single-vector models face theoretical capacity limits~\cite{weller2026theoreticallimitationsembeddingbasedretrieval}, researchers have increasingly turned to multi-vector architectures. Pioneered in text by ColBERT~\cite{khattab2020colbertefficienteffectivepassage} and ColBERTv2~\cite{santhanam2022colbertv2}, late-interaction mechanisms have been adapted for multimodal tasks via models like XTR~\cite{lee2023xtr}, Colpali~\cite{faysse2025colpali}, jina-embeddings-v4~\cite{gunther2025jinaembeddingsv4universalembeddingsmultimodal}, and MetaEmbed~\cite{xiao2026metaembedscalingmultimodalretrieval}. Unlike these approaches%
, which all learn a \emph{dedicated} token-level representation and require full-scale task-specific training, adapters, or learnable tokens%
, SMART can read late-interaction tokens directly off the hidden states of a \emph{frozen} single-vector model and add them to that model's own pooled score, allowing it to be used entirely inference-only. %

\paragraph{Reusing Frozen Representations}
Another line of work also reuses or recombines frozen representations, but differs from SMART in mechanism. Late Chunking~\cite{gunther2024latechunking} reuses frozen token states but mean-pools them, using neither \textsc{MaxSim} nor a hybrid score; HEAVEN~\cite{kim2026heaven} combines single- and multi-vector retrieval as a two-stage cascade rather than as a single fused score; and two concurrent preprints recover different signals, with Spectral Retrieval~\cite{morandi2026spectral} interpolating \textsc{MaxSim} and mean-pooling through a sinc kernel, and Latent Terms~\cite{clavie2026latentterms} extracting a sparse lexical BM25 signal. To our knowledge, no prior work holds SMART's conjunction: recovering ColBERT-style dense late interaction, optionally training-free, from a frozen single-vector multimodal embedder, fused additively with that model's own pooled score.

\paragraph{Multimodal Retrieval Benchmarks}
Standardizing evaluation has evolved from foundational baselines like M-BEIR~\cite{wei2023uniirtrainingbenchmarkinguniversal} to comprehensive collections like MMEB~\cite{jiang2025vlm2vectrainingvisionlanguagemodels} and MMEB-V2~\cite{meng2025vlm2vecv2advancingmultimodalembedding}, which span diverse modalities and tasks. Targeted benchmarks have also emerged for specific domains, including ViDoRe~\cite{faysse2025colpali} and VisRAG~\cite{yu2024visrag} for visual documents, Jina-VDR~\cite{gunther2025jinaembeddingsv4universalembeddingsmultimodal} for image retrieval, and UMRB~\cite{zhang2025gmeimprovinguniversalmultimodal} for unified retrieval. In this work, we evaluate SMART on MMEB-V2 due to its broad inclusion of dense retrieval tasks across image, document, and video domains.

\section{SMART}

In this section, we present SMART, which stands for \textbf{S}ingle-to-\textbf{M}ulti \textbf{A}daptation for \textbf{R}etrieval \textbf{T}ransformers. We first provide some preliminaries over existing single-vector embedders and their limitations in Sec.~\ref{sec: method prelim}. We then dive into the observation that led to the design of SMART in Sec.~\ref{sec: method main}. Lastly, we analyze conditions of applying SMART in Appendix~\ref{sec: task cat}.

\subsection{Preliminaries: Single-vector Objective and Bottleneck}
\label{sec: method prelim}

Multimodal embedding models are typically built on rich token-level encoders~\cite{li2026qwen3vlembeddingqwen3vlrerankerunifiedframework, meng2025vlm2vecv2advancingmultimodalembedding}, but they are trained and used through a much narrower readout. Given an input, the encoder produces a sequence of hidden states over text tokens, visual tokens, and special tokens. In standard contrastive training, however, supervision is applied only to a designated pooling representation, most commonly the final-layer hidden state of the end-of-text (\texttt{eot}) token. For a query $q$, a positive candidate $c^{+}$, and a set of negatives $\{c^{-}\}$, the model is optimized with the InfoNCE loss~\cite{oord2018infonce}:

\vspace{-1em}
\begin{equation}
    \mathcal{L}
    =
    -\log
    \frac{
        \exp\!\bigl(s_{\mathrm{single}}(q, c^{+}) / \tau\bigr)
    }{
        \exp\!\bigl(s_{\mathrm{single}}(q, c^{+}) / \tau\bigr)
        +
        \sum_{c^{-}} \exp\!\bigl(s_{\mathrm{single}}(q, c^{-}) / \tau\bigr)
    } ,
    \label{eq:contrastive}
\end{equation}

where the score is computed from the normalized \texttt{eot} representations:

\vspace{-1em}
\begin{equation}
    s_{\mathrm{single}}(q, c)
    =
    \left(h^{L}_{q,\texttt{eot}}\right)^{\top}
    h^{L}_{c,\texttt{eot}}  . 
\end{equation}

Thus, although the encoder maintains a full sequence of token-level representations, the training signal directly supervises only the pooled embedding. At retrieval time, the same single-vector readout is used, so each query and candidate is collapsed into one normalized embedding, and ranking reduces to nearest-neighbor search in a shared embedding space.

Single-vector retrieval achieves efficiency by compressing any input into one pooled embedding. This compression induces the \emph{single-vector bottleneck}, where that single representation must support the entire retrieval decision even when relevance heavily depends on localized evidence. This is especially pronounced in fine-grained multimodal retrieval, where details confined to a small portion of the candidate (text or image) are crucial. Consequently, a high single-vector similarity score may indicate aggregate semantic relatedness while completely ignoring localized information.
Prior late-interaction and multi-vector retrievers~\cite{khattab2020colbertefficienteffectivepassage,santhanam2022colbertv2, faysse2025colpali,xiao2026metaembedscalingmultimodalretrieval} alleviate this limitation by retaining token- or patch-level representations and computing relevance through local interactions. 
These methods, however, typically require full-scale training, incurring substantial computational and memory costs as self-attention cost grows quadratically with sequence length~\cite{vaswani2017attentionisallyouneed}. This motivates the question: \emph{can we extend an existing single-vector retriever with multi-vector capabilities while preserving its original backbone and efficient pooled representation?}

\subsection{Direct Late Interaction over Hidden States}
\label{sec: method main}

\paragraph{Pooled supervision reaches non-pooling hidden states}
To approach this question, we examine the supervision dynamics of contrastive retrieval training. At first glance, the contrastive loss in Eq.~\eqref{eq:contrastive} appears to supervise only the pooled embeddings, suggesting that contrastive training mainly shapes the pooling token. This interpretation overlooks the fact that the pooled state is a function of the full token sequence. Through the transformer's attention and residual pathways, $h^{L}_{q,\texttt{eot}}$ aggregates information from every non-pooling token, so any token that contributes to the pooled state lies on the gradient path of the contrastive loss: 
\begin{equation}
    \frac{\partial \mathcal{L}}{\partial h^{l}_{q,i}}
    =
    \left(
    \frac{\partial z_q}
         {\partial h^{l}_{q,i}}
    \right)^{\!\top}
    \frac{\partial \mathcal{L}}
         {\partial z_q},
    \label{eq:gradient-flow}
\end{equation}
where $h^{l}_{q,i}$ denotes the hidden state of the $i$-th query token at layer $l$, $L$ is the final layer, and $z_q$ is the normalized pooled embedding. %
The pooled contrastive objective optimizes the global representation for cosine-based matching, while its gradients propagate through the token-level computation that produces it. This motivates us to examine whether the resulting non-pooling hidden states also support cosine-based retrieval, despite receiving no explicit token-level supervision.

\paragraph{Pooled supervision leaves hidden states usable for cosine matching}
We compare Qwen3-VL-Embedding-2B with its generative initialization, Qwen3-VL-2B-Instruct~\cite{bai2025qwen3-vltechnicalreport}. The former receives embedding-specific training through a single global representation, whereas the latter is trained only with generative objectives such as next-token prediction. To isolate the retrieval signal in the non-pooling states, we score both models using \emph{only} \textsc{MaxSim} over their final-layer non-pooling hidden states, without the global term.
Embedding-2B reaches 61.2 Hit@1 on image retrieval and 68.7 NDCG@5 on visual documents, against 15.6 and 7.9 for Instruct-2B. This shows that the hidden states of the embedding model support cosine-based late interaction while those of the generative model largely do not. Since the embedding objective directly supervises only the single global representation, this provides empirical evidence that such training also shapes the non-pooling hidden states into a geometry useful for cosine matching. This is consistent with \citet{chen2021intriguing}, who show that instance-level contrastive learning can produce meaningful local and hierarchical features even when its objective operates only on a global image representation.

\paragraph{Single-to-Multi Adaptation for Retrieval Transformers}

Motivated by this, we propose SMART, a single-to-multi adaptation that reuses the hidden states of a single-vector retriever for additional token-level retrieval.
In its most basic form, this adaptation can be applied even without training a new multi-vector retriever. We keep the original backbone and pooled readout, and add a token-level late-interaction readout over the hidden states already produced by the model.

Importantly, we use this token-level signal as a complement to the original pooled score, not as a replacement. The pooled score captures global query-candidate compatibility, while token-level matching can expose local evidence that may be compressed away by the pooled readout. To combine these two signals without an additional projection or rescaling step, we use final-layer non-pooling hidden states for the token-level readout. We use the final layer rather than earlier layers because the pooled embedding is read out from this layer, making it most directly compatible with the original single-vector scoring space. Note that this is not a claim that earlier layers lack useful information, as they may encode rich lexical, visual, and local details%
; Appendix~\ref{sec: layer appendix} shows the most useful information is carried across a broad mid-to-late region, and the best late-interaction layer is backbone-dependent.

Let $M_q$ and $M_c$ denote the valid non-pooling token indices of query $q$ and candidate $c$, respectively, excluding padding tokens and the pooling token. For each token, we use the normalized final-layer hidden state $\tilde{h}^{L}_{x,i}=h^{L}_{x,i}/\|h^{L}_{x,i}\|_2$. We compute a \textsc{MaxSim} late-interaction score~\cite{khattab2020colbertefficienteffectivepassage}
by matching each query token to its most similar candidate token:
\begin{equation}
    s_{\mathrm{late}}(q,c)
    =
    \frac{1}{|M_q|}
    \sum_{i \in M_q}
    \max_{j \in M_c}
    \tilde{h}_{q,i}^{L\,\top}
    \tilde{h}_{c,j}^{L}.
    \label{eq:maxsim}
\end{equation}
The late-interaction score measures local query coverage in the candidate hidden states. Since it is computed in the same final-layer cosine geometry as the pooled readout, SMART combines it with the original single-vector score by simple addition:
\begin{equation}
    s_{\mathrm{hybrid}}(q,c)
    =
    s_{\mathrm{single}}(q,c)
    +
    s_{\mathrm{late}}(q,c).
    \label{eq:hybrid-score}
\end{equation}
We use unit weighting to keep SMART hyperparameter-free (Section~\ref{sec: lambda} verifies this choice). %
A candidate ranks highly under $s_{\mathrm{hybrid}}$ when it is both globally compatible with the query and locally supported by token-level evidence. While SMART can be applied at inference time without any training, we also explore using $s_{\mathrm{hybrid}}$ as the training objective in Appendix~\ref{sec: ablate hybrid}, where we demonstrate how training with hybrid scoring provides the most performance gain.

\section{Experiments}

In this section, we conduct experiments and analyses using SMART in both inference-only and training scenarios. We first use a controlled experiment to validate our hypothesis of using local evidence for retrieval in Section~\ref{sec:toy-dataset}. We then show inference-only results and the sensitivity of hybrid scoring to its weighting in Section~\ref{sec: inf results}, results of training a SMART adapter in Section~\ref{sec: adapter}, and results of training and converting our own models in Section~\ref{sec: lora}. We then conduct qualitative analysis over some visualizations of SMART in Section~\ref{sec: qual}. %
We additionally report the per-layer analysis in Appendix~\ref{sec: layer appendix} and  deployment cost in  Appendix~\ref{sec: efficiency}.

\subsection{Controlled Local-Evidence Toy Benchmark}
\label{sec:toy-dataset}

\begin{figure}[t]
    \centering
    \includegraphics[width=\textwidth]{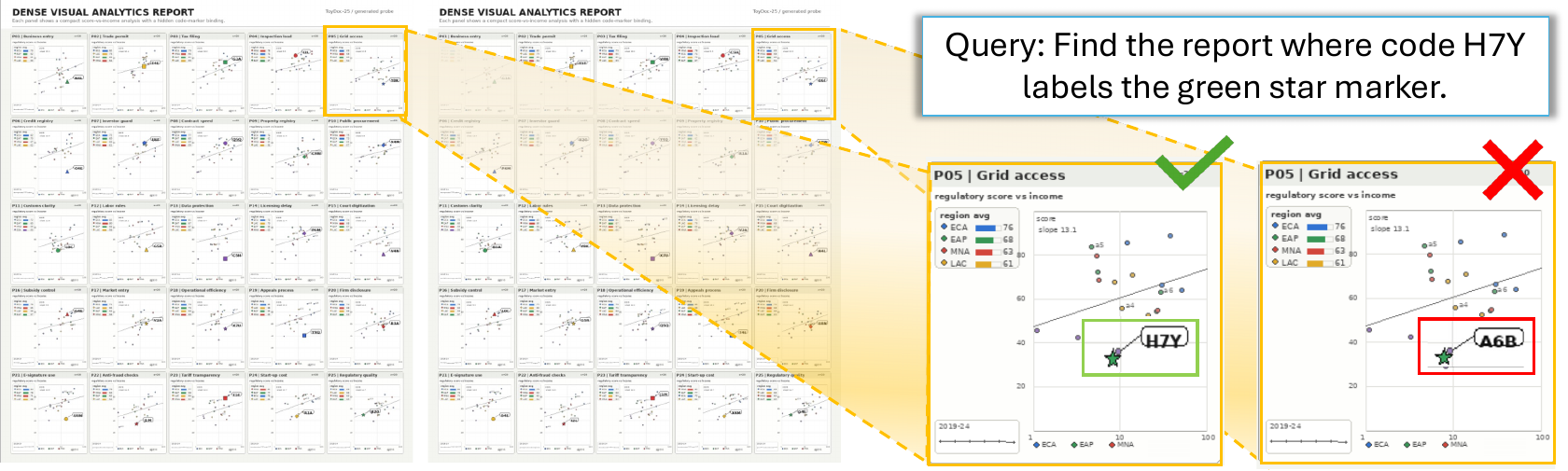}
    \vspace{-1em}
    \caption{
    Controlled local-evidence toy benchmark. Each query specifies a local
    code--marker binding. The hard negative has the same layout, codes, colors, and shapes as the positive report, but the codes are reassigned so that no code remains paired with its original marker.
    }
    \label{fig:toy-dataset}
\end{figure}

To make the local-evidence bottleneck of pooled single-vector retrieval explicit, we construct a controlled pairwise benchmark over dense visual reports. This deliberately adversarial setting limits the usefulness of global similarity, allowing us to examine whether retrievers can distinguish reports using local evidence. As shown in Figure~\ref{fig:toy-dataset}, each example consists of a positive report $d_A$ and a hard negative $d_B$, both rendered as a $5 \times 5$ grid of chart panels. Each panel contains one local binding between an alphanumeric code and a visual marker described by color and shape. %
The hard negative keeps the same layout, codes, and markers but permutes their pairings, so success depends on recognizing the local binding rather than detecting whether either element appears (Appendix~\ref{sec: toy dataset appendx}).

We generate $40$ report pairs with $25$ bindings per pair, yielding $1000$ queries. We first evaluate the setting with full overlap ($o=1.00$) of codes and marker descriptors. Each query ranks only its corresponding positive and hard-negative reports, and we report pairwise accuracy. %
As shown in Table~\ref{tab:toy-overlap}, the original single-vector score selects the positive report %
well below chance, showing that the pooled single-vector readout is unreliable when relevance is determined by a specific local binding. Replacing the single score with a late-interaction score over final-layer non-pooling hidden states %
improves accuracy substantially, showing that these hidden states expose local code--marker binding evidence that is not reliably accessible through the pooled readout.
The gap between these scores suggests that the bottleneck lies in the pooled single-vector readout rather than in the absence of local information in the model. %
This difficulty is not peculiar to single-vector models. In the same pairwise setting at $o=1.00$, jina-embeddings-v4 in its native multi-vector mode and Colpali achieve accuracies of $50.9\%$ and $48.7\%$, respectively. Both score near chance and below late interaction over Qwen3-VL-Embedding-2B's frozen hidden states, suggesting that local code--marker binding remains difficult even for retrievers explicitly trained for token-level matching.

\begin{table}[t]
    \centering
    \caption{Toy benchmark pairwise accuracy (\%) as code and marker overlap $o$ varies (chance $=50\%$).}
    \label{tab:toy-overlap}
    \small
    \begin{tabular}{lcccccc}
        \toprule
        Overlap $o$ & 0.04 & 0.20 & 0.40 & 0.60 & 0.80 & 1.00 \\
        \midrule
        $s_{\mathrm{single}}$          & \textbf{70.0} & 67.0 & 61.0 & 52.0 & 49.0 & 31.9 \\
        $s_{\mathrm{late}}$            & 68.5 & 66.0 & 62.0 & 60.0 & \textbf{55.0} & \textbf{56.8} \\
        $s_{\mathrm{hybrid}}$ (ours)   & 69.6 & \textbf{68.2} & \textbf{65.0} & \textbf{62.0} & 52.0 & 42.6 \\
        \bottomrule
    \end{tabular}
\end{table}

\paragraph{When does the pooled term help?}
We then reduce the overlap $o$ between the positive and negative reports. The negative retains a fraction $o$ of the positive report's codes and marker descriptors, while the remainder are replaced with items from a disjoint pool. At intermediate overlap ($o=0.20$--$0.60$), hybrid scoring outperforms both pooled and late-interaction scoring (Table~\ref{tab:toy-overlap}). These results suggest that global similarity can complement local evidence when reduced overlap provides additional cues for distinguishing the reports. Since overlap is a property of the corpus rather than a parameter chosen at deployment, we also examine how closely natural corpora approach this controlled setting. Essentially no real document pairs reach the benchmark's near-duplicate regime (Appendix~\ref{sec: toy dataset appendx}).

\subsection{Inference-Only Results: SMART's Plug-and-Play Effectiveness}
\label{sec: inf results}

\begin{table}[t]
    \centering
    \caption{Average per-task performance on MMEB-V2. We report Recall@1 for image and video retrieval tasks and NDCG@5 for visual document retrieval tasks. Applying training-free SMART achieves consistent improvement %
on nearly all dense retrieval tasks and generalizes across different backbones, including the SoTA Qwen3-VL-Embedding series.}
    \label{tab:task_types}
    \begin{tabular}{l|c cccc c|c} 
        \toprule
        \multirow{2.5}{*}{\textbf{Model}} & \textbf{Image} & \multicolumn{4}{c}{\textbf{Visdoc}} & \textbf{Video} & \multirow{4}{*}{\textbf{Average}} \\
        
        \cmidrule(lr){2-2} \cmidrule(lr){3-6} \cmidrule(lr){7-7}
        
        & RET & VDRv1 & VDRv2 & VR & OOD & RET & \\
        \cmidrule(lr){1-7}
        \# Tasks & 12 & 10 & 4 & 6 & 4 & 5 & \\
        \midrule
        
        VLM2Vec-V2.0~\cite{meng2025vlm2vecv2advancingmultimodalembedding}             & 69.50 & 75.34 & 47.28 & 79.43 & 62.04 & 28.66 & 64.50 \\
        \quad\quad\quad\quad\quad\quad \textbf{+ SMART}          & \textbf{69.95} & \textbf{80.23} & \textbf{51.08} & \textbf{82.76} & \textbf{64.00} & \textbf{30.03} & \textbf{67.04} \\
        GME-2B~\cite{zhang2025gmeimprovinguniversalmultimodal}              & 70.13 & 87.61 & 53.76 & 82.49 & 66.93 & \color{gray}{26.73} & 69.00 \\
        \quad\quad\quad\quad\quad\quad \textbf{+ SMART}          & \textbf{71.05} & \textbf{87.97} & \textbf{57.23} & \textbf{84.18} & \textbf{67.82} & \color{gray}{26.50} & \textbf{70.00} \\
        GME-7B~\cite{zhang2025gmeimprovinguniversalmultimodal} & 73.09 & 90.01 & 60.43 & 86.20 & 69.22 & \color{gray}{29.92} & 72.26 \\
        \quad\quad\quad\quad\quad\quad \textbf{+ SMART} &\textbf{73.57} & \textbf{90.12} & \textbf{61.04}& \textbf{87.05} & \textbf{69.43} & \color{gray}{29.38} & \textbf{72.56}\\
        Qwen3-VL-Embed-2B~\cite{li2026qwen3vlembeddingqwen3vlrerankerunifiedframework} & 74.91 & 84.60 & 65.33 & 86.34 & 69.27 & 54.04 & 74.92 \\
        \quad\quad\quad\quad\quad\quad \textbf{+ SMART}          & \textbf{75.33} & \textbf{85.52} & \textbf{66.61} & \textbf{86.87} & \textbf{69.90} & \textbf{56.05} & \textbf{75.77} \\
        Qwen3-VL-Embed-8B~\cite{li2026qwen3vlembeddingqwen3vlrerankerunifiedframework} & 80.09 & 87.29 & 69.35 & 88.78 & 73.27 & 59.01 & 78.83 \\
        \quad\quad\quad\quad\quad\quad \textbf{+ SMART}          & \textbf{80.15} & \textbf{87.92} & \textbf{70.57} & \textbf{89.12} & 73.21 & \textbf{60.43} & \textbf{79.34} \\
        \bottomrule
    \end{tabular}
\end{table}

Table \ref{tab:task_types} presents the comprehensive evaluation of inference-only SMART across dense retrieval tasks within the MMEB-V2~\cite{meng2025vlm2vecv2advancingmultimodalembedding} benchmark (Apache-2.0 License). We note reasons for selecting these tasks in Appendix~\ref{sec: task cat}. %
The improvements brought forth by SMART are consistent across retrieval domains and backbones. All of them are obtained inference-only, with no parameter updates. %

\textbf{%
Compatibility Across Backbones.} An important feature of SMART is its %
generalizability across different models.
SMART %
improves baselines like VLM2Vec-V2.0, driving an overall average improvement of +2.54\%.
Furthermore, SMART's efficacy is not limited to weaker baselines; it %
also applies to SoTA architectures. When applied to the %
Qwen3-VL-Embedding series, SMART still %
yields consistent gains. On Qwen3-VL-Embedding-2B, we observe %
a +0.86-point average improvement. Even on the larger Qwen3-VL-Embedding-8B, SMART %
raises the average from 78.83\% to 79.34\%.
A paired $t$-test over the full 24-task visual-document suite confirms that the gains on both Qwen3-VL-Embedding backbones are significant at $p < 10^{-4}$ (Appendix~\ref{sec: significance}).

\paragraph{Sensitivity of hybrid scoring}
\label{sec: lambda}
Eq.~\eqref{eq:hybrid-score} combines the two scores by unweighted addition, so it is worth asking how much rests on that choice. Table~\ref{tab:lambda-sweep} reports a sweep over $\lambda$ on the same two inference-only suites, where $\lambda = 0$ and $\lambda = 1$ recover the single-vector and adding \mbox{SMART} entries of Table~\ref{tab:task_types}. Ranking quality is flat across $\lambda \in [0.5, 2]$ on both suites and degrades only at the extremes: at $\lambda = 4$ the local term overwhelms and late interaction alone falls well below the single-vector baseline on both suites. %
The unweighted sum is therefore a robust default rather than a tuned hyperparameter, and the two terms are genuinely complementary: neither alone matches their combination.

\begin{table}[t]
    \centering
    \caption{Sensitivity of retrieval quality to the weight $\lambda$ in $s = s_{\mathrm{single}} + \lambda \cdot s_{\mathrm{late}}$, on Qwen3-VL-Embedding-2B. NDCG@5 and Hit@1 are reported for Visdoc and image retrieval, respectively. %
}
    \label{tab:lambda-sweep}
    \begin{tabular}{l|cccccc}
        \toprule
        $\lambda$ & single-only & 0.5 & 1 & 2 & 4 & late-only \\
        \midrule
        Visdoc (24 tasks)   & 79.27 & 79.82 & 80.10 & \textbf{80.20} & 79.59 & 68.70 \\
        Image RET (12 tasks) & 74.91 & 75.26 & \textbf{75.33} & 74.89 & 73.84 & 61.19 \\
        \bottomrule
    \end{tabular}
\end{table}

\textbf{Robustness Across Retrieval Domains.} The granular task breakdown further highlights SMART's versatility. In the complex domain of Visual Document Retrieval (Visdoc; VDRv1, VDRv2, VR, and OOD subsets), where fine-grained text-to-visual alignment is particularly important, the addition of SMART demonstrates consistent improvements across all four tested backbones.
Similarly, in Video Retrieval, SMART %
improves VLM2Vec (+1.37\%), Qwen3-VL-Embedding-2B (+2.01\%), and Qwen3-VL-Embedding-8B (+1.42\%). We gray out GME because it is not trained to handle multiple frames, and thus when only the middle frame is provided (as was done in MMEB-V2~\cite{meng2025vlm2vecv2advancingmultimodalembedding}), we see minimal change. We further analyze SMART's performance concentration and tradeoffs in Appendix~\ref{sec: coarse fine}.

In summary, the empirical evidence %
supports SMART as a training-free enhancement that consistently improves dense retrieval across multimodal backbones. Its use of a per-token index and MaxSim scoring introduces additional storage and inference costs. We report the storage requirements, index size, query latency, and reranking depth in Appendix~\ref{sec: efficiency}.

\begin{table}[t]
    \centering
    \caption{Results of training SMART adapters when evaluated on MMEB-V2's visdoc subset. In the SMART columns, \cmark denotes SMART hybrid scoring $s_{\mathrm{hybrid}}$. \xmark$_{\tiny s}$ denotes no SMART with single-vector scoring only, while \xmark$_{\tiny m}$ denotes no SMART with late-interaction/multi-vector scoring only. \cmark$^\dagger$ means freezing the model and only training an $s_{\mathrm{late}}$ adapter.}
    \label{tab:visdoc_adapter}
    \resizebox{\linewidth}{!}{\begin{tabular}{l|c|cc|cccc|c}
        \toprule
        \multirow{2.5}{*}{\textbf{Model}} 
        & \multirow{2.5}{*}{\textbf{Size}} 
        & \multicolumn{2}{c|}{\textbf{SMART}} 
        & \multicolumn{4}{c|}{\textbf{Visdoc}} 
        & \multirow{2.5}{*}{\textbf{Average}} \\
        \cmidrule(lr){3-8}
        & & Train & Eval & VDRv1 & VDRv2 & VR & OOD & \\
        \midrule
        \multicolumn{9}{l}{\textit{SoTA Single-Vector Embedders (Qwen3-VL-Embedding Family)}}\\
        \midrule
        \multirow{6.5}{*}{Qwen3-VL-Embedding}
        & \multirow{3}{*}{2B} & \xmark$_{\tiny s}$ & \xmark$_{\tiny s}$
        & 84.60 & 65.33 & 86.34 & 69.27 & 79.27 \\
        
        &  & \xmark$_{\tiny s}$ & \cmark 
        & 85.52 & 66.61 & 86.87 & 69.90 & 80.10 \\
        
        &  & \cmark$^{\dagger}$ & \cmark 
        & \textbf{87.09} & \textbf{67.08} & \textbf{87.99} & \textbf{70.73} & \textbf{81.25} \\
        \cmidrule{2-9}
        & \multirow{3}{*}{8B} & \xmark$_{\tiny s}$ & \xmark$_{\tiny s}$ 
        & 87.29 & 69.35 & 88.78  & 73.27 & 82.33 \\
        
        &  & \xmark$_{\tiny s}$ & \cmark 
        & 87.92 & 70.57 & 89.12  & 73.21 & 82.88 \\
        
        &  & \cmark$^{\dagger}$ & \cmark 
        & \textbf{89.42} & \textbf{71.25} & \textbf{89.67} & \textbf{73.99} & \textbf{83.89} \\

        \midrule
        \multicolumn{9}{l}{\textit{SoTA Multi-Vector Embedders}}\\
        \midrule
        Colpali-1.3~\cite{faysse2025colpali} 
        & 3B & \xmark$_{\tiny m}$ & \xmark$_{\tiny m}$ 
        & 83.60 & 52.00 & 81.10 & 43.10 & 71.00\\
        
        jina-embeddings-v4~\cite{gunther2025jinaembeddingsv4universalembeddingsmultimodal} 
        & 4B & \xmark$_{\tiny m}$ & \xmark$_{\tiny m}$
        & 89.94 & 57.36 & 88.74 & 70.18 & 80.91\\
        \bottomrule
    \end{tabular}}
\end{table}

\subsection{Lightweight Adapter Post-Training}
\label{sec: adapter}

We next investigate whether late interaction can benefit from a minimally learned readout. %
To isolate %
its effect, we keep the embedder frozen and train only a token-wise linear adapter on top of the final-layer hidden states.
For each valid hidden state $h_i^L \in \mathbb{R}^{H}$, the adapter applies layer normalization followed by a linear projection and $\ell_2$ normalization:
\begin{equation}
    r_i
    =
    \mathrm{normalize}\!\left(
        \mathrm{Linear}\!\left(\mathrm{LN}(h_i^L)\right)
    \right),
    \label{eq:linear-adapter}
\end{equation}
where $\mathrm{Linear}: \mathbb{R}^{H}\rightarrow\mathbb{R}^{d}$ is the only
trainable readout module. We use the Colpali~\cite{faysse2025colpali} training set with global batch size 512. Training is very efficient as the adapter for Qwen3-VL-Embedding-2B only takes 1 hour and 50 minutes on one node of eight 48GB A6000s. We replace the normalized hidden states in Eq.~\eqref{eq:maxsim} with the adapted token vectors $r_i$ while keeping the pooled single-vector score unchanged and apply the same hybrid scoring as Eq.~\eqref{eq:hybrid-score}. The adapter is trained only with $s_{\mathrm{late}}$.

This readout-only adapter improves over the training-free SMART variant across both model sizes, as demonstrated in Table~\ref{tab:visdoc_adapter}. We see %
a 1-point gain on both the 2B and 8B Qwen3-VL-Embedding variants%
. The consistent gain suggests that the frozen backbone already contains local evidence, and that a lightweight token-level readout can make this evidence more compatible with late interaction.

Most importantly, %
\textbf{Qwen3-VL-Embedding-2B becomes competitive with the SoTA multi-vector model jina-embeddings-v4} with the help of the SMART adapter, edging ahead by 0.34 points with roughly half the parameters (2B vs.\ 4B) after under two hours of training on academia-level resources; Appendix~\ref{sec: seeds} confirms this is not a single favorable seed. That margin is small and across models, so it leaves open whether SMART merely approximates a trained multi-vector head. jina-embeddings-v4 settles this within one model, as it exposes both a single-vector and a trained native multi-vector mode. We apply \emph{inference-only} SMART to each over the 24-task visual-document suite that is jina-embeddings-v4's home domain. In single-vector mode, inference-only SMART improves NDCG@5 from $76.29$ to $81.44$ without additional training, exceeding the $80.91$ achieved by jina's trained multi-vector head (Table~\ref{tab:visdoc_adapter}). Applying SMART to the trained multi-vector head yields a marginal improvement, from $80.91$ to $80.95$. These results suggest that SMART complements single-vector models and can reduce the need for dedicated multi-vector training, without implying that purpose-trained multi-vector models are inferior. %

\subsection{Efficient Conversion via LoRA Finetuning with SMART}
\label{sec: lora}

One may ask, ``Why would one not simply train a multi-vector model from scratch for better performance?'' %
We answer this by training our own single- and multi-vector embedders.
 We adopt the LamRA-Ret recipe~\cite{liu2024lamralargemultimodalmodel} to finetune the backbone Qwen3-VL-2B-Instruct~\cite{bai2025qwen3-vltechnicalreport} (the same starting model as Qwen3-VL-Embedding-2B).
 All training is done on one node of eight 80GB A100s with global batch size 512, using LoRA~\cite{hu2021loralowrankadaptationlarge} with $r=128$, $\alpha=256$, cosine annealing scheduling with max learning rate $1e-4$ and a warmup ratio of 0.03. We only train on the Colpali~\cite{faysse2025colpali} training set for 4 epochs for LamRA-Single and LamRA-Multi, while for LamRA-Single-Convert we trained on top of LamRA-Single for only one more epoch.
 
 The results are shown in Table~\ref{tab:visdoc_lamra}. The first model, LamRA-Single, is trained only using the $s_\mathrm{single}$ objective and takes 6.5 hours. When we compare the baseline single-vector model to the inference-only application of SMART (Rows 1 and 2), we observe a solid $\sim$1.6-point average improvement (72.60 to 74.18), consistent with the training-free boosts established in Section~\ref{sec: inf results}.

In Row 3, we start from the LamRA-Single checkpoint, and apply the exact same recipe but only train for one more epoch, this time using the $s_\mathrm{hybrid}$ objective, and take an extra 3 hours (9.5 hours in total) to train LamRA-Single-Convert. We also train another model, LamRA-Multi, from the Qwen3-VL-2B-Instruct model, only using the $s_{late}$ objective, which takes 12 hours. %
Both outperform the Single variants, but the converted model takes significantly shorter to train in total ($\sim20\%$) while performing only slightly behind the Multi variant ($\sim0.63$). Thus, our answer to the question is: \textbf{Using SMART to convert a single-vector model into a multi-vector model is more efficient than training one from scratch without significant loss of performance.}

\begin{table}[t]
    \centering
    \caption{Results of training and converting single-vector models when evaluated on the visdoc subset. \xmark$_{\tiny s}$ denotes no SMART with single-vector scoring only, while \xmark$_{\tiny m}$ denotes no SMART with late-interaction/multi-vector scoring only. \cmark denotes SMART hybrid scoring $s_{\mathrm{hybrid}}$ and \cmark$^\dagger$ means extended training (\textit{not} from scratch) with LoRA using $s_{\mathrm{hybrid}}$. }
    \label{tab:visdoc_lamra}
    \resizebox{\linewidth}{!}{\begin{tabular}{l|c|cc|cccc|c}
        \toprule
        \multirow{2.5}{*}{\textbf{Model}}  & \multirow{2.5}{*}{\textbf{Training Time}} 
        & \multicolumn{2}{c|}{\textbf{SMART}} 
        & \multicolumn{4}{c|}{\textbf{Visdoc}} 
        & \multirow{2.5}{*}{\textbf{Average}} \\
        \cmidrule(lr){3-8}
        & & Train & Eval & VDRv1 & VDRv2 & VR & OOD & \\
        \midrule

        \multicolumn{8}{l}{\textit{Our Trained Embedders (LamRA-Ret Family)}}\\
        \midrule
        LamRA-Single  & 6.5 hours & \xmark$_{\tiny s}$ & \xmark$_{\tiny s}$ 
        & 81.58 & 50.72 & 78.41 & 63.50 & 72.60\\
        
        LamRA-Single-SMART 
         & 6.5 hours& \xmark$_{\tiny s}$ & \cmark 
        & 83.02 & 52.25 & 80.52 & 64.50 & 74.18\\

        LamRA-Single-Convert 
        & \textbf{9.5 hours}& \cmark$^\dagger$ & \cmark 
        & 86.93 & 54.60 & 84.39 & 67.61 & 77.68\\
        
        LamRA-Multi 
         & 12 hours & \xmark$_{\tiny m}$ & \xmark$_{\tiny m}$
        & 87.93 & 54.29 & 85.24 & 67.91 & 78.31\\
        \bottomrule
    \end{tabular}}
\end{table}

\subsection{Qualitative Analysis}
\label{sec: qual}

\begin{figure}[t]
    \centering
    \includegraphics[width=\linewidth]{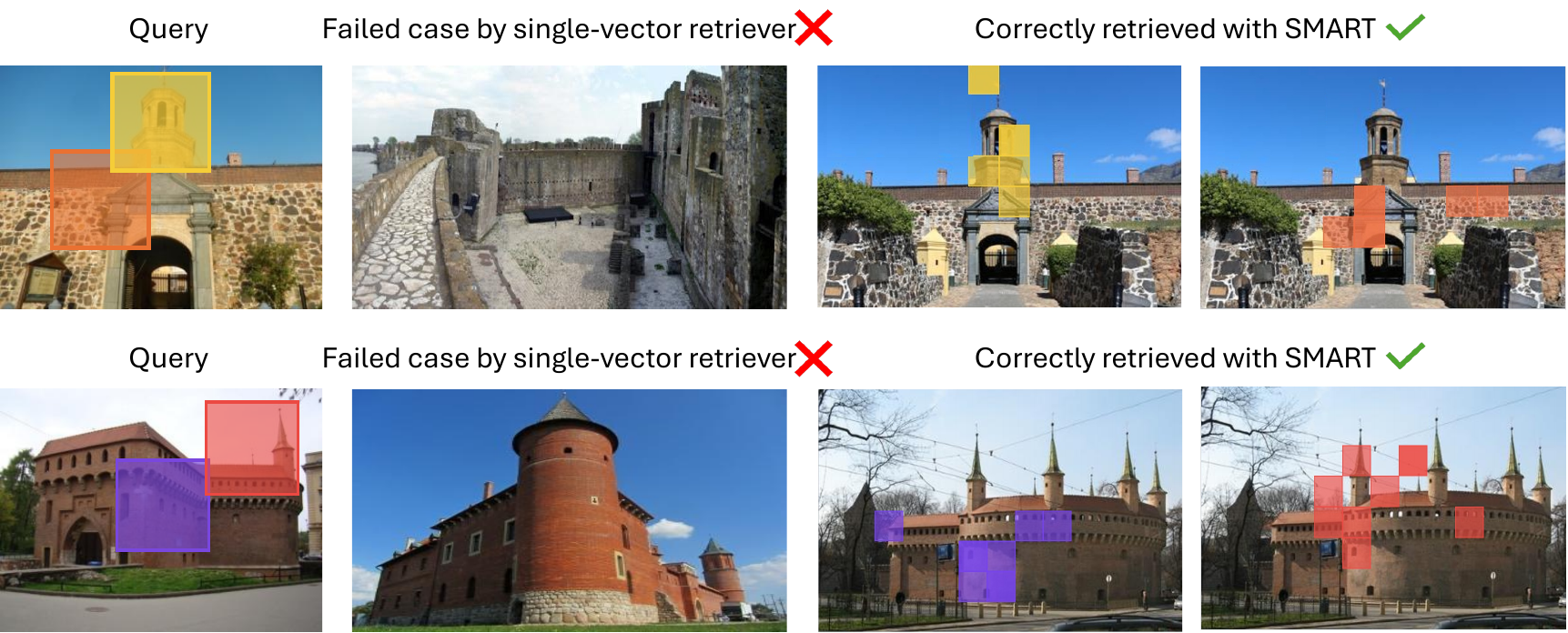}
    \caption{Qualitative visualization of SMART on image-to-image retrieval. Each row shows the query, a failed result from the original single-vector retriever, and the correct result retrieved by SMART. Colored boxes visualize the local evidence used by SMART. Each selected query patch is matched to the highest-similarity candidate patch under the \textsc{MaxSim} late-interaction score, with matching colors denoting the corresponding query-candidate pair. These localized matches help SMART recover details that are often obscured by global single-vector compression.}
    \label{fig:token_vis}
\end{figure}

SMART is designed to complement the global signal of a single-vector retriever with token-level evidence from hidden states. We qualitatively examine this complementarity on examples in Figures~\ref{fig:teaser} and~\ref{fig:token_vis} to illustrate how the pooled score can retrieve a globally similar but incorrect candidate, while the late-interaction signal helps recover local visual details that support the correct match. We include more visualization analyses in Appendix~\ref{appendix:vis}.

Figure~\ref{fig:teaser} demonstrates a scenario where a SoTA multimodal retrieval model, Qwen3-VL-Embedding-8B, fails to retrieve one of three relevant images for a query in the Vidore Economic Reports task (from the MMEB-V2 Visual Document subset). However, the application of SMART successfully corrects this failure. A closer examination of the visual content reveals fine-grained details within a chart legend (specifically, the labels ``Europe and Central Asia'' and ``Middle East and North Africa'') which align perfectly with the query text. This highlights a fundamental limitation of single-vector models: they often fail to capture granular details because they compress the entire document into a single, global representation optimized for broad comparisons across diverse queries and corpora. In contrast, SMART effectively recovers this localized information.

Furthermore, Figure~\ref{fig:token_vis} shows that the original single-vector retriever can select globally plausible but incorrect candidates. These errors are natural under a pooled representation, as the retrieved images often share the same broad visual category, such as castles, fortresses, towers, or stone architecture, but differ in the specific local structures needed to identify the correct instance. The visualization further shows where SMART obtains its token-level evidence. For selected query tokens, we highlight the candidate-image token with the highest hidden-state cosine similarity, which is exactly the token selected by the \textsc{MaxSim} operation in $s_{\mathrm{late}}$. These highlighted regions concentrate on small, semantically meaningful parts of the candidate image rather than spreading uniformly, indicating that individual query tokens match specific local regions instead of averaged global content. Such localized cues can be obscured by the pooled readout when the entire candidate is summarized into a single embedding.

\section{Conclusion}

In this work, we introduced SMART (Single-to-Multi Adaptation for Retrieval Transformers) to overcome the fundamental information bottleneck inherent in single-vector multimodal retrieval models. By %
showing empirically that contrastive training on a global pooling token %
leaves the preceding hidden states usable for %
cosine-based matching, we %
unlocked these localized representations using a late-interaction \textsc{MaxSim} operator and our special hybrid scoring objective.

Our extensive experiments highlight SMART's dual utility as both an effective inference-time enhancement and an efficient training paradigm. In an inference-only setting, SMART functions as a zero-cost, plug-and-play upgrade that consistently improves accuracy across diverse modalities, scaling robustly to SoTA architectures like the Qwen3-VL-Embedding series. Crucially, when training only lightweight auxiliaries, SMART further improves performance by explicitly optimizing the non-pooling-token hidden states for late interaction together with the global summary token, allowing us to convert a SoTA single-vector embedding model into a multi-vector variant efficiently.

Ultimately, SMART provides a robust, computationally efficient pathway for recovering fine-grained, localized evidence from existing multimodal architectures, raising the ceiling for universal dense retrieval systems.

\section*{Limitations}
This work focuses on dense retrieval tasks, as we find SMART not beneficial when used as an inference-only tool for more global tasks like classification. Due to limited compute, we could only train our LamRA-Ret models on the visdoc subset. SMART's main remaining cost is a per-token index (Appendix~\ref{sec: efficiency}). Its gains are also concentrated, largest on visual documents and absent on models that have a trained multi-vector head (Section~\ref{sec: adapter}). We leave the exploration of these aspects as valuable future work.

\newpage

\section*{Acknowledgments}
This work was supported in part by NSF IIS2404180, NetApp Inc., Institute of Information \& communications Technology Planning\& Evaluation (IITP) grants (MSIT) (No. 2022-0-00871, Development of AI Autonomy and Knowledge Enhancement for AI Agent Collaboration), (No. RS-2025-2543949. Environment-Aware and Domain-Adaptive Multimodal Embodied AI for Real-World Interaction), the Artificial Intelligence Graduate School Program at Korea University (No. RS-2019-II190079) and grant No. RS-2025-25439490, the National Research Foundation of Korea (NRF) grant funded by the Korea government (MSIT) (RS-2025-25302986); and Culture, Sports and Tourism R\&D Program through the Korea Creative Content Agency grant funded by the Ministry of Culture, Sports and Tourism in 2024 (Project Name: International Collaborative Research and Global Talent Development for the Development of Copyright Management and Protection Technologies for Generative AI, Project Number: RS-2024-00345025).

{
    \small
    \bibliographystyle{ieeenat_fullname}
    \bibliography{main}
}

\appendix

\section{Applicable Task Categorization}
\label{sec: task cat}
In this work, we focus our evaluation on dense, corpus-level retrieval tasks (e.g., Image, Visual Document, and Video Retrieval) where the semantic complexity of the query and target necessitates fine-grained alignment. While SoTA single-vector models compress the entire multi-modal representation into a designated token (e.g., \texttt{<eot>}), this creates an information bottleneck for complex retrieval. Conversely, SMART leverages late interaction across all hidden states between the query and the corpus, preserving rich, token-level semantics. 

We explicitly distinguish these complex retrieval tasks from classification, standard VQA, visual grounding, and temporal localization tasks (particularly in MMEB-V2~\cite{meng2025vlm2vecv2advancingmultimodalembedding} that we report on), as they present fundamentally different architectural and definitional challenges:

    \paragraph{Classification and VQA:} In these tasks, the target is typically a low-entropy concept (e.g., ``dog'' or ``43'') that is easily compressed into a single vector. Applying SMART here may offer limited benefit. Furthermore, forcing dense, token-level interaction on single-label tasks may introduce noise, as the model attempts to find complex local alignments that may be less relevant to the target label.
    
    \paragraph{Visual Grounding (Image):} We omit image-based visual grounding due to inherent ambiguities when cast as an open-corpus retrieval problem. In standard grounding, the objective is to localize a specific crop within a \textit{provided} source image. When expanded to a retrieval formulation across a broader corpus, a query for ``dog'' becomes confounded by valid distractor crops from \textit{other} images. The model is penalized for retrieving semantically correct matches due to these spurious correlations, which does not match the purpose of dense corpus-level retrieval.
    
    \paragraph{Video Moment Retrieval:} Finally, while video moment retrieval features a better-defined corpus (e.g., candidate clips isolated within a specific movie), the required semantic granularity from considering groups of tokens can pose challenges for SMART. Identifying a dynamic action like ``running'' requires holistic spatiotemporal abstraction. SMART computes fine-grained, token-by-token similarities and, in its inference-only configuration, does not add explicit temporal reasoning to align high-level, abstract action semantics to singular spatial patches within individual frames. Future work could explore intermediate trainable modules that aggregate temporal continuous features into discrete semantic units, making them amenable to SMART.

\paragraph{Slight Architectural Adjustments for Composed Image Retrieval}
For Composed Image Retrieval (CIR) benchmarks, spurious correlations \cite{zhang2026reasoningaugmentedrepresentationsmultimodalretrieval} persist and negatively affect model performance. 
Thus, at inference time, we modify SMART for these tasks by masking the query vision tokens to reduce misleading visual alignments and improve overall retrieval accuracy.

\section[Layer-wise Analysis Details]{Layer-wise Analysis Details}
\label{sec: layer appendix}

We examine which layer should supply the late-interaction tokens, holding the pooled anchor fixed at the final layer as the training objective defines it. Table~\ref{tab:layer-backbones} reports the sweep. On the full 24-task suite, Qwen3-VL-Embedding-2B varies by less than $0.8$ points across layers 4 to 28, and layer 20 ($80.16$) is only marginally better than the final layer ($80.10$), so the final layer is not uniquely optimal. The best-performing layer also varies by backbone. jina-embeddings-v4 achieves its highest score at its final layer ($89.40$), ahead of every intermediate layer by more than 2.5 points, whereas both Qwen3-VL-Embedding models peak mid-to-late, the 8B model at layer 28 of 36 and the 2B model at layer 20 of 28. Useful local evidence is therefore distributed across a broad mid-to-late region whose exact optimum depends on the backbone, making the final layer a simple default. %

\begin{table}[!tbp]
    \centering
    \caption{Layer sweep with the pooled anchor held at the final layer; only the layer $L$ supplying the late-interaction tokens varies. \textit{Top block}: the complete 10-task ViDoRe\_v1 benchmark across three backbones, showing that the best layer is \emph{not} universal. \textit{Bottom block}: the same sweep for Qwen3-VL-Embedding-2B over the full 24-task visual-document suite, showing relatively stable hybrid performance across the tested token layers. }
    \label{tab:layer-backbones}
    \resizebox{\textwidth}{!}{
    \begin{tabular}{l|ccccccccc}
        \toprule
        Late-interaction layer $L$ & 4 & 8 & 12 & 16 & 20 & 24 & 28 & 32 & 36 \\
        \midrule
        \multicolumn{10}{l}{\textit{ViDoRe\_v1 (10 tasks), NDCG@5}} \\
        Qwen3-VL-Embedding-2B (28 layers) & 84.90 & 85.20 & 85.08 & 85.06 & \textbf{85.56} & 85.31 & 85.52 & --- & --- \\
        jina-embeddings-v4 (36 layers)    & 85.59 & 86.61 & 86.81 & 86.88 & 86.87 & 86.79 & 86.32 & 86.74 & \textbf{89.40} \\
        Qwen3-VL-Embedding-8B (36 layers) & 87.55 & 88.00 & 87.93 & 87.82 & 87.84 & 88.00 & \textbf{88.16} & 87.95 & 87.92 \\
        \midrule
        \multicolumn{10}{l}{\textit{Full 24-task visual-document suite, NDCG@5}} \\
        Qwen3-VL-Embedding-2B (28 layers) & 79.42 & 79.83 & 79.67 & 79.70 & \textbf{80.16} & 80.04 & 80.10 & --- & --- \\
        \bottomrule
    \end{tabular}
    }
\vspace{1em}
    \centering
    \caption{Layer-wise analysis on visual-document retrieval tasks. We compare using the same layer $X$ for both pooled and token-level readouts with keeping the original last-layer pooled score fixed while varying only the hidden layer $X$ used for late interaction.}
    \label{tab:visdoc_layers_pooling}
    \resizebox{\textwidth}{!}{
    \begin{tabular}{l||cccc|c||cccc|c}
        \toprule
        \multirow{2.5}{*}{\textbf{$X$}}
        & \multicolumn{5}{c||}{\textbf{$X$-th Layer Pooling}}
        & \multicolumn{5}{c}{\textbf{Last Layer Pooling}} \\
        \cmidrule(lr){2-6} \cmidrule(lr){7-11}
        & VDRv1 & VDRv2 & VR & OOD & Average
        & VDRv1 & VDRv2 & VR & OOD & Average \\
        \midrule
        4  & 56.77 & 25.53 & 57.93 & 37.18 & 48.59 & 84.90 & 64.94 & 86.66 & 69.31 & 79.42 \\
        8  & 72.94 & 39.25 & 69.87 & 47.28 & 62.28 & 85.20 & 66.01 & 87.00 & 69.44 & 79.83 \\
        12 & 75.04 & 42.40 & 70.44 & 52.94 & 64.77 & 85.08 & 65.64 & 86.76 & 69.51 & 79.67 \\
        16 & 71.46 & 40.22 & 63.30 & 50.43 & 60.71 & 85.06 & 65.90 & 86.70 & 69.63 & 79.70 \\
        20 & 80.45 & 55.01 & 78.34 & 60.40 & 72.34 & \textbf{85.56} & 66.42 & \textbf{87.19} & 69.87 & \textbf{80.16} \\
        24 & 82.68 & 60.93 & 83.41 & 64.99 & 76.29 & 85.31 & 66.41 & 87.16 & 69.80 & 80.04 \\
        \midrule
        28 & \textbf{85.52} & \textbf{66.61} & \textbf{86.87} & \textbf{69.90} & \textbf{80.10} & 85.52 & \textbf{66.61} & 86.87 & \textbf{69.90} & 80.10 \\
        \bottomrule
    \end{tabular}
    }
\end{table}

Table~\ref{tab:visdoc_layers_pooling} gives the per-subset breakdown, together with a complementary configuration. On the right, the pooled anchor is held at the final layer and only the layer $X$ supplying the hidden states varies; the Average column of that block is the bottom row of Table~\ref{tab:layer-backbones}. On the left, layer $X$ supplies \emph{both} the pooled vector and the hidden states. The left block degrades sharply for early $X$ ($48.59$ at $X = 4$, against $79.42$ for the same hidden states anchored at the final layer), which supports using the final-layer pooled representation while allowing earlier hidden states to supply local evidence. This is why we keep the pooled anchor at the final layer throughout.

\section{Ablation for Hybrid Scoring}
\label{sec: ablate hybrid}

Table~\ref{tab:visdoc_hybrid} compares the LamRA scoring configurations. LamRA-Hybrid combines trained \textsc{MaxSim} interaction with the pooled single-token score and achieves the highest average NDCG@5 among these configurations (79.10), a 6.5-point improvement over LamRA-Single.

LamRA-Hybrid also approaches the average NDCG@5 of Qwen3-VL-Embedding-2B (79.10 versus 79.27). These results suggest that hybrid scoring can support retrieval performance when used during training, in addition to its inference-only application.

\begin{table}[H]
    \centering
    \caption{Ablation of training and evaluation scores for the LamRA-Ret family on the MMEB-V2 visual-document suite (NDCG@5). \xmark$_{\tiny s}$ denotes no SMART with single-vector scoring only, while \xmark$_{\tiny m}$ denotes no SMART with late-interaction/multi-vector scoring only. \cmark denotes SMART hybrid scoring $s_{\mathrm{hybrid}}$.}
    \label{tab:visdoc_hybrid}
    \resizebox{\linewidth}{!}{\begin{tabular}{l|c|cc|cccc|c}
        \toprule
        \multirow{2.5}{*}{\textbf{Model}} 
        & \multirow{2.5}{*}{\textbf{Size}} 
        & \multicolumn{2}{c|}{\textbf{SMART}} 
        & \multicolumn{4}{c|}{\textbf{Visdoc}} 
        & \multirow{2.5}{*}{\textbf{Average}} \\
        \cmidrule(lr){3-8}
        & & Train & Eval & VDRv1 & VDRv2 & VR & OOD & \\
        \midrule

        \multicolumn{9}{l}{\textit{Our Trained Embedders (LamRA-Ret Family)}}\\
        \midrule
        LamRA-Single 
        & \multirow{4.5}{*}{2B} & \xmark$_{\tiny s}$ & \xmark$_{\tiny s}$ 
        & 81.58 & 50.72 & 78.41 & 63.50 & 72.60\\
        
        LamRA-Single-SMART 
        &  & \xmark$_{\tiny s}$ & \cmark 
        & 83.02 & 52.25 & 80.52 & 64.50 & 74.18\\
        
        LamRA-Multi 
        & & \xmark$_{\tiny m}$ & \xmark$_{\tiny m}$ 
        & 87.93 & 54.29 & 85.24 & 67.91 & 78.31\\
        
        LamRA-Hybrid 
        & & \cmark & \cmark 
        & \textbf{88.14} & \textbf{55.81} & \textbf{86.39} & \textbf{68.90} & \textbf{79.10}\\
        \bottomrule
    \end{tabular}}
    \vspace{-1em}
\end{table}

\section[Deployment Cost: Storage, Latency, and Reranking Depth]{Deployment Cost: Storage, Latency, and Reranking Depth}
\label{sec: efficiency}

Inference-only SMART requires no additional training, but its token-level index and MaxSim scoring incur serving costs. We report these costs below.

\paragraph{Candidate storage and index size}
By default, SMART stores the full 2048-d hidden state for every document token. This is tunable: a lightweight \emph{unsupervised} PCA projection, fit in seconds on the corpus tokens with no labels and no gradient training, reduces each token to 128-d, matching the token dimension of jina-embeddings-v4's native multi-vector index. Table~\ref{tab:storage} gives the per-document cost. The projection is applied post hoc to the frozen hidden states used by SMART.

\begin{table}[htbp]
    \centering
    \caption{Serving cost of SMART: candidate storage per document, and scoring latency per query on one RTX A6000 (fp16, 2048-d) against a pool of $C = 100$ candidates. SMART-PCA uses an unsupervised PCA projection to reduce the token dimension from 2048 to 128. SMART adds no forward pass, so the latency shown is MaxSim scoring only; we did not separately profile scoring at 128-d. Storage is an offline resource and standard post-hoc compression applies on top, but it remains the main cost SMART introduces relative to single-vector retrieval.}
    \label{tab:storage}
    \begin{tabular}{lccc}
        \toprule
        Index configuration & Bytes / doc & vs.\ single & Scoring latency, $C=100$ \\
        \midrule
        Single-vector (2048-d)                        & 4 KB    & $1\times$    & $\sim$0.03 ms \\
        SMART-PCA (128d)        & 184 KB  & $46\times$   & --- \\
        SMART, full 2048-d                            & 2.88 MB & $737\times$  & $\sim$3--5 ms \\

        \bottomrule
    \end{tabular}
\end{table}

Table~\ref{tab:pca-matched} evaluates PCA compression on an 18-task subset of the visual-document suite (the full 24 minus the 6 largest-corpus sets, omitted for computational cost). With 128-d token vectors, SMART over jina-embeddings-v4's frozen single-vector states scores above its trained native multi-vector mode (83.98 versus 83.19) and achieves comparable performance to full 2048-d SMART (83.89), with a $16\times$ reduction in token dimension.

\begin{table}[htbp]
    \centering
    \caption{Comparison of 128-d and 2048-d token representations on an 18-task subset of the visual-document suite (NDCG@5). SMART-PCA achieves a higher score than jina-embeddings-v4's native 128-d multi-vector head and a comparable score to full 2048-d SMART.}
    \label{tab:pca-matched}
    \begin{tabular}{lc}
        \toprule
        18-task subset & NDCG@5 \\
        \midrule
        jina native multi-vector (trained, 128-d)             & 83.19 \\
        jina-single + SMART-PCA (128-d)  & \textbf{83.98} \\
        jina-single + SMART, full 2048-d                      & 83.89 \\
        \bottomrule
    \end{tabular}
\end{table}

\paragraph{Query latency}
SMART reuses the hidden states from the pooled-vector computation without an additional backbone forward pass. Table~\ref{tab:storage} reports its MaxSim scoring cost. For full 2048-d SMART with $C=100$ candidates, processing $100$ queries as a batch yields an average MaxSim scoring cost of $0.13$ ms per query. For fixed query and candidate token counts, the amount of MaxSim computation scales linearly with $C$. These timings cover scoring only and exclude query encoding.

\paragraph{Reranking depth}
The main experiments score each query against every candidate in the corpus. Table~\ref{tab:rerank} evaluates an alternative that retrieves the top-100 candidates using $s_{\mathrm{single}}$ and reranks them with the hybrid score. This approach matches the reported average score of exhaustive scoring at the displayed precision (75.33). It limits MaxSim scoring to 100 candidates per query, corresponding to the $C = 100$ setting in Table~\ref{tab:storage}. Reranking the same candidates using $s_{\mathrm{late}}$ alone scores below the single-vector baseline, suggesting that the pooled term remains useful in this setting.

\begin{table}[htbp]
    \centering
    \caption{Top-100 reranking on the 12-task MMEB-V2 benchmark (Qwen3-VL-Embedding-2B, Recall@1). Hybrid reranking matches the reported average score of exhaustive scoring at the displayed precision, while reranking by $s_{\mathrm{late}}$ alone yields a lower score.}
    \label{tab:rerank}
    \begin{tabular}{lc}
        \toprule
        Scheme & Recall@1 \\
        \midrule
        Single-vector                                                        & 74.91 \\
        Full-corpus hybrid ($\lambda = 1$, ours)                             & \textbf{75.33} \\
        Retrieve top-100 by $s_{\mathrm{single}}$ $\rightarrow$ rerank by hybrid       & \textbf{75.33} \\
        Retrieve top-100 by $s_{\mathrm{single}}$ $\rightarrow$ rerank by pure $s_{\mathrm{late}}$ & 61.6 \\
        \bottomrule
    \end{tabular}
\end{table}

\section{Toy Dataset}
\label{sec: toy dataset appendx}

The pooling bottleneck is difficult to isolate in natural retrieval benchmarks, where global semantics, local evidence, and dataset biases are often entangled. We therefore construct a controlled toy benchmark that directly tests whether a retriever can distinguish local bindings from aggregate content. The goal of this probe is not to model real documents in full generality, but to isolate retrieval decisions that depend on local code--marker bindings. Each example consists of a pair of dense visual documents, denoted by $(d_A,d_B)$. Both documents have the same layout: a $5 \times 5$ grid of panels, where each panel contains a cluttered chart and one local code--marker binding. A marker is described by a color and a shape, and is labeled by a short alphanumeric code. For each pair, $d_A$ is the positive document. The hard negative $d_B$ preserves the same layout, the same set of codes, and the same set of visual marker descriptors, but permutes the code assignments across panels. Thus, for every query, the negative document contains the queried code and the queried visual descriptor, but not in the correct local binding. This construction removes easy global cues. Since $d_A$ and $d_B$ contain the same objects, colors, shapes, codes, and document structure, a model cannot solve the task by checking whether the requested elements are present somewhere in the page. It must instead determine whether the code and the visual marker are bound to each other in the same local region. This is precisely the kind of evidence that can be obscured by a single pooled embedding.

We generate $40$ document pairs. Each pair contains $25$ code--marker bindings,
and we create one query for each binding, yielding $1000$ queries in total. A query asks for a specific local binding, e.g., ``find the report where code $x$ labels the red star marker.'' For each query, we rank only the two documents from the same pair: the positive document $d_A$ and the hard negative $d_B$. We report pairwise accuracy, the fraction of queries for which the scoring function assigns a higher score to $d_A$ than to $d_B$.

Table~\ref{tab:toy-overlap} reports pairwise accuracy across code and marker overlap levels, including the full-overlap setting ($o=1.00$) used here. Late interaction alone outperforms the pooled score, indicating that non-pooling hidden states retain local binding evidence, while the hybrid score improves over the pooled score but remains below late interaction alone, because this adversarial benchmark intentionally makes the pooled global signal misleading. Section~\ref{sec:toy-dataset} examines the effect of reducing overlap, with hybrid scoring outperforming both pooled and late-interaction scoring at $o=0.20$--$0.60$.

To assess how this controlled setting compares with natural data, we sampled $50$k random document pairs per corpus from the MMEB-V2 visual-document suite. These pairs match on $28.4\%$ of pixels on average (median $31.2\%$), and only $0.035\%$ reach at least $99\%$ pixel agreement. Near-identical document pairs are therefore rare among these randomly sampled pairs. The toy benchmark serves as a controlled probe of local binding evidence under deliberately high similarity, rather than a representative retrieval task.

\section[Statistical Significance]{Statistical Significance}
\label{sec: significance}

The inference-only results in Table~\ref{tab:task_types} show improvements across backbones, with small absolute margins on the strongest models. We assess these gains against test-set sampling variation.

For inference-only SMART there is no training and no stochastic component: the model parameters and the evaluation procedure are both fixed, so repeated evaluations of the same checkpoint do not measure variation from training. Estimating checkpoint-level variability instead would require independently trained versions of Qwen3-VL-Embedding-2B and 8B, which are not publicly available. What we can test is whether the observed gain is large relative to the variation induced by the choice of evaluation queries. We therefore run a paired $t$-test on the per-query NDCG@5 differences (single vs.\ hybrid) over the full 24-task visual-document suite, with $N = 12{,}590$ queries per backbone, and obtain 95\% confidence intervals on the per-query mean gain by bootstrap.

\begin{table}[htbp]
    \centering
    \caption{Statistical significance of the inference-only SMART gain on the full 24-task visual-document suite. Paired $t$-test on per-query NDCG@5 differences (single vs.\ hybrid), $N = 12{,}590$ queries per backbone; 95\% confidence intervals on the per-query mean gain obtained by bootstrap. Both intervals exclude zero.}
    \label{tab:significance}
    \begin{tabular}{lcccc}
        \toprule
        Backbone (24-task visdoc suite) & Gain (NDCG@5) & Paired $t$ & $p$ & Bootstrap 95\% CI \\
        \midrule
        Qwen3-VL-Embedding-2B & $+0.83$ & 10.2 & $< 10^{-4}$ & $[+0.67, +0.99]$ \\
        Qwen3-VL-Embedding-8B & $+0.48$ & 5.09 & $< 10^{-4}$ & $[+0.25, +0.56]$ \\
        \bottomrule
    \end{tabular}
\vspace{-1em}
\end{table}

Table~\ref{tab:significance} reports the results. On both backbones, the paired test gives $p < 10^{-4}$ and the bootstrap confidence interval excludes zero, supporting a positive average gain on the evaluated query set despite the modest absolute improvement.

The gains in Table~\ref{tab:significance} are averaged over \emph{queries}, whereas Tables~\ref{tab:task_types} and~\ref{tab:visdoc_adapter} report \emph{task}-level means, so the two need not agree exactly. For Qwen3-VL-Embedding-8B the query-averaged gain is $+0.48$ while the task-averaged gain is $82.33 \rightarrow 82.88 = +0.55$. Both describe the same underlying improvement under different weightings of the suite.

\vspace{-1em}
\section[Coarse- versus Fine-Grained Tasks]{Coarse- versus Fine-Grained Tasks}
\label{sec: coarse fine}
\vspace{-1em}

Table~\ref{tab:coarse-fine} compares the effect of inference-only SMART across MMEB-V2 image tasks. Image classification changes only slightly, from $70.29$ to $70.01$. The reported retrieval gains are larger on document- and composition-heavy tasks, while performance on coarse whole-image retrieval is approximately unchanged. This pattern is consistent with token-level matching being more useful when retrieval depends on fine-grained evidence. When late interaction is not beneficial for a task, setting $\lambda = 0$ recovers the original single-vector score.

\vspace{-1em}
\begin{table}[H]
    \centering
    \caption{Coarse- versus fine-grained trade-off for inference-only SMART on MMEB-V2 image tasks (Qwen3-VL-Embedding-2B, Recall@1 gains in percentage points). Classification performance changes only slightly, while the reported retrieval gains are larger on document- and composition-heavy tasks. Setting $\lambda = 0$ recovers the original single-vector score.}
    \label{tab:coarse-fine}
    \begin{tabular}{@{}lr@{}}
        \toprule
        Task(s) & SMART gain (points) \\
        \midrule
        Image classification (10 tasks) & $-0.28$ \\
        \midrule
        Wiki-SS-NQ (document screenshots) & $+1.5$ \\
        WebQA & $+1.5$ \\
        CIRR (composed retrieval) & $+1.9$ \\
        MSCOCO, VisualNews (coarse whole-image) & $\approx$ neutral \\
        \bottomrule
    \end{tabular}
\vspace{-1em}
\end{table}

\section[Adapter Seed Variability]{Adapter Seed Variability}
\label{sec: seeds}
\vspace{-1em}
Evaluations across multiple backbones demonstrate generality but do not substitute for repeated trials when training is involved. For the lightweight-adapter experiments of Section~\ref{sec: adapter} we therefore trained an additional adapter for each of Qwen3-VL-Embedding-2B and 8B under a different random seed, keeping the training configuration otherwise unchanged. Table~\ref{tab:seeds} reports both runs. The additional-seed run closely reproduces the main-paper result in each case: the overall averages differ by 0.09 points for the 2B model and 0.06 points for the 8B model, with two-run standard deviations of 0.07 and 0.04 respectively. This suggests that the reported adapter performance is not driven by a single favorable seed.
\vspace{-1em}
\begin{table}[H]
    \centering
    \caption{Adapter training repeated under a second random seed, evaluated on the MMEB-V2 visual-document subset.}
    \label{tab:seeds}
    \begin{tabular}{l|cccc|c}
        \toprule
        Run & VDRv1 & VDRv2 & VR & OOD & Average \\
        \midrule
        \multicolumn{6}{l}{\textit{Qwen3-VL-Embedding-2B}} \\
        Main-paper run        & 87.09 & 67.08 & 87.99 & 70.73 & 81.25 \\
        Additional-seed run   & 86.90 & 67.40 & 87.74 & 70.68 & 81.16 \\
        \midrule
        \multicolumn{6}{l}{\textit{Qwen3-VL-Embedding-8B}} \\
        Main-paper run        & 89.42 & 71.25 & 89.67 & 73.99 & 83.89 \\
        Additional-seed run   & 89.63 & 71.02 & 89.65 & 74.10 & 83.95 \\
        \bottomrule
    \end{tabular}
\end{table}

\section{Visualization}
\label{appendix:vis}

\begin{figure}[ht]
    \centering
    \includegraphics[width=\linewidth]{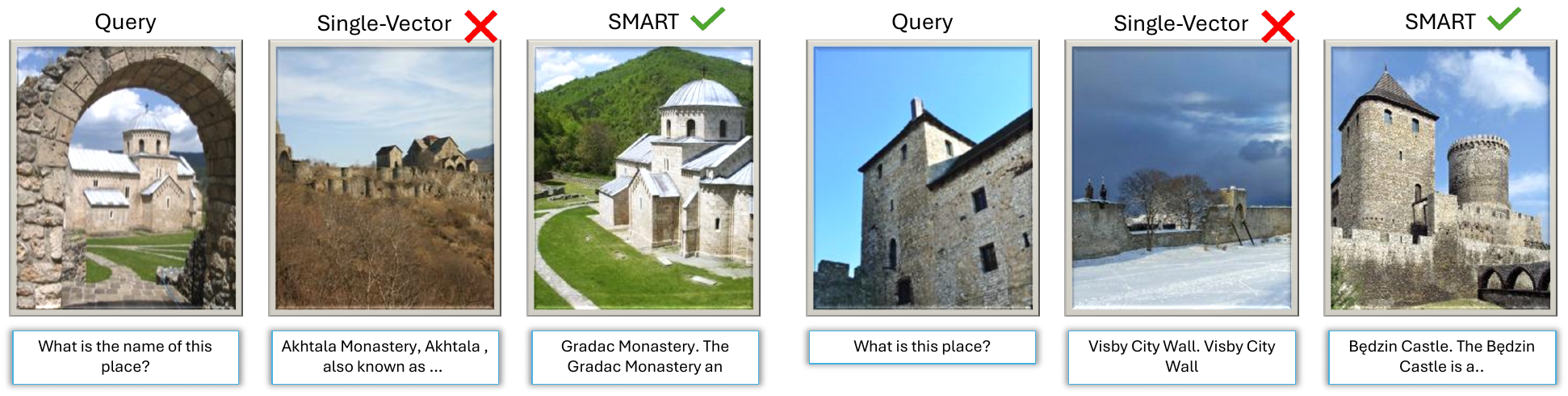}
    \caption{
    Additional qualitative examples where the original single-vector retriever fails but SMART retrieves the correct candidate. The single-vector model can select images that are globally plausible but miss localized visual evidence, while SMART improves retrieval by adding token-level matching over hidden states.
    }
    \label{fig:appendix-failed-cases}
\end{figure}

\begin{figure}[ht]
    \centering
    \includegraphics[width=\linewidth]{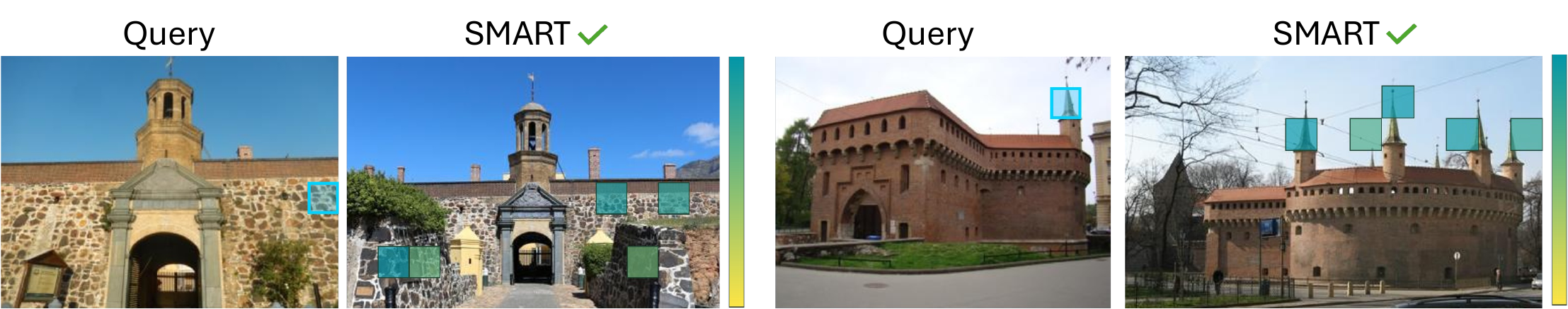}
    \caption{
    Token-level visualization of SMART. For selected query image tokens, we show
    the top-$5$ candidate-image tokens. The highlighted regions illustrate that SMART's late-interaction score captures localized visual evidence rather than only global image similarity.
    }
    \label{fig:appendix-single-token}
\end{figure}

Figure~\ref{fig:appendix-failed-cases} shows additional cases where the original single-vector retriever fails but SMART retrieves the correct candidate. These examples illustrate a limitation of relying only on a pooled representation. The single-vector result is often globally plausible, sharing broad visual semantics such as stone buildings, castles, towers, or monastery-like structures, but can miss localized details needed to identify the correct instance. SMART corrects these cases by complementing the pooled score with late interaction over the remaining hidden states, allowing local visual evidence in the query to be matched against corresponding regions in the candidate image.

Figure~\ref{fig:appendix-single-token} further visualizes where this token-level evidence comes from. For each selected query token, we highlight the top-$5$ candidate-image tokens with the highest hidden-state cosine similarity. The highest-scoring token corresponds to the match used by the \textsc{MaxSim} aggregation in $s_{\mathrm{late}}$, while the remaining high-similarity tokens reveal nearby local evidence supporting the same token-level comparison. The highlighted patches concentrate on semantically meaningful regions, such as towers, roof structures, walls, entrances, and distinctive architectural details, rather than spreading uniformly over the image. This suggests that SMART recovers localized correspondences that can be obscured when the entire image is compressed into a single global embedding.

\end{document}